\documentclass[12pt,superscriptaddress,floatfix,a4paper]{article}
\usepackage[T2A]{fontenc}
\usepackage[cp1251]{inputenc}
\usepackage[english]{babel}
\usepackage{amsmath,amssymb,amsfonts}
\usepackage{bm,latexsym,euscript,textcomp}
\usepackage{graphicx,color,graphics,caption,epsf,dcolumn}
\usepackage{natbib}
\usepackage{hyperref}
\usepackage{epstopdf}
\newcommand{\beq}{\begin{equation}}
\newcommand{\eeq}{\end{equation}}
\newcommand{\pt}{\partial}

\begin{document}

\title{\Large \bf Why does air passage over forest yield more rain?\\
Examining the coupling between rainfall, pressure and atmospheric moisture content}

\author{A. M. Makarieva$^1$\thanks{\textit{Corresponding author.} {E-mail: ammakarieva@gmail.com}}, V. G. Gorshkov$^1$, D.  Sheil$^{2,3,4}$,
A. D. Nobre$^5$,\\ P. Bunyard$^6$, B.-L. Li$^7$}

\date{\vspace{-5ex}}

\maketitle

$^1$Theoretical Physics Division, Petersburg Nuclear Physics Institute, 188300, Gatchina,
St. Petersburg, Russia; $^2$School of Environment, Science and Engineering, Southern Cross University, PO Box 157, Lismore, NSW 2480, Australia;
$^3$Institute of Tropical Forest Conservation, Mbarara University of Science and Technology, PO Box, 44, Kabale, Uganda;
$^4$Center for International Forestry Research, PO Box 0113 BOCBD, Bogor 16000, Indonesia;
$^5$Centro de Ci\^{e}ncia do Sistema Terrestre INPE, S\~{a}o Jos\'{e} dos Campos SP 12227-010, Brazil;
$^6$Lawellen Farm, Withiel, Bodmin, Cornwall, PL30 5NW, United Kingdom and University Sergio Arboleda, Bogota, Colombia;
$^7$XIEG-UCR International Center for Arid Land Ecology, University of California, Riverside 92521-0124, USA.

\begin{abstract}
The influence of forest loss on rainfall remains poorly understood. Addressing this challenge
Spracklen et al. recently presented a pan-tropical study of rainfall and land-cover that showed that satellite-derived rainfall measures
 were positively correlated with the degree to which model-derived air trajectories had been exposed to forest cover. This result confirms
 the influence of vegetation on regional rainfall patterns suggested in previous studies. However, we find that the conclusion
 of Spracklen et al. -- that differences in rainfall reflect air moisture content resulting from evapotranspiration
while the circulation pattern remains unchanged -- appears undermined by methodological inconsistencies.
We identify methodological problems with the underlying analyses and the quantitative estimates for rainfall change
predicted if forest cover is lost in the Amazon.
We discuss some alternative explanations that include the distinct role
 of forest evapotranspiration in creating low pressure systems that draw moisture from the oceans to the continental hinterland.
 Our analysis of meteorological data from three regions in Brazil, including the central Amazon forest, reveal
a tendency for rainy days during the wet season
with column water vapor (CWV) exceeding 50 mm to have higher pressure than rainless days; while at lower CWV rainy days
tend to have lower pressure than rainless days. The coupling between atmospheric moisture content
and circulation dynamics  underlines that the danger posed by forest loss is greater than suggested by focusing only on moisture recycling alone.
\end{abstract}

\section{Introduction}

The ongoing loss of natural forest cover in many regions has caused many concerns.  These concerns include the resulting changes in local and regional rainfall patterns and their reliability.  Many of the key relationships are contested but valuable new approaches and data are increasingly available.  In their recent pan-tropical study \citet{spr} examined how air exposure to forest cover influences subsequent rainfall
from air moving over the tropical land surface.  The positive association found confirms the influence of vegetation on
regional rainfall patterns suggested in previous studies \citep{mg07,doug09,ch10,goe11,cook11,mgl09,mgl13,du12}.
The approach taken by \citet{spr} -- reconstruction of the air trajectories from
the observed wind fields and the use of daily rainfall statistics -- allowed them
to analyze the apparent short-term influence of the forest
on rainfall. Such an approach offers new insights into the vegetation-rainfall relationship but the guiding concepts for the emerging
interpretations require scrutiny.

In this paper we revisit the arguments that led
\citet{spr} to conclude that the additional rain reflects a higher air moisture content resulting from forest evapotranspiration.
In Section 2 we show that the quantitative conclusion about the importance of evapotranspiration
varies with respect to the chosen time length of the air trajectories and the period of the recorded rainfall.
In Section 3 we show that Spracklen et al.'s estimate of post-deforestation reduction in Amazonian rainfall is itself a function of the
selected duration of the air trajectories.
We discuss the implications of such arbitrary choices for the interpretations offered by \citet{spr}.

A key assumption underlying the analysis of \citet{spr} was that the atmospheric circulation remains unaffected by the presence or absence of
forests, such that forests only impact rainfall by modifying the amount of moisture in the atmosphere via evapotranspiration.
In Section 4 we discuss the physical variables that determine rainfall and
present evidence for the tight dynamic coupling between
rainfall, atmospheric moisture content and such parameters of atmospheric circulation
like atmospheric pressure and wind direction. We show that the results of \citet{spr}
of a comparatively higher rainfall produced by forest compared to non-forest air are consistent with the interpretation that
that large forests create low-pressure systems that may bring rain to the adjacent areas,
while the air arrived from non-vegetated areas should be more often associated with high pressure, descending air motion and little rain.
In the concluding section we discuss the key features of forests as a regional rainmaking system.

\section{Additional rainfall and evapotranspiration}

\citet{spr} used the Tropical Rainfall Measurement Mission (TRMM) data in the form of daily rainfall
corresponding to $1^{\rm o} \times 1^{\rm o}$ degree cells for the 2001-2007 time period. They matched each local rainfall measurement to
a trajectory that described the air's motion during the ten days preceding
the measurement (Fig.~\ref{spr1}). Air trajectories were calculated with the OFFLINE trajectory model \citep{methven97}.
Using satellite-derived monthly mean leaf area index (LAI)
\citep{myneni02}, \citet{spr} quantified the air's exposure to forest by integrating the leaf area index (LAI) of the vegetation
traversed over the preceding 10-days.

This magnitude, denoted $\Sigma {\rm LAI}$, is measured in time units (days), as LAI itself is dimensionless:
\beq
\label{sumLAI}
\Sigma {\rm LAI} = \overline {\rm LAI} \times t_L, \,\,\,t_L \equiv \frac{L}{u} \le t_T = 10\, {\rm d}.
\eeq
Here $\overline {\rm LAI}$ is the mean leaf area index on the traversed territory,
$t_L$ is the duration of the terrestrial part of the trajectory by the air moving with velocity $u$,
$L$ is length of the terrestrial part of the trajectory, $t_T$ is the total duration of the trajectory (including
its oceanic part not shown in Fig.~\ref{spr1}). Note that thus defined $\Sigma {\rm LAI}$ depends on $t_L$.

For each of the four tropical regions they studied, \citet{spr} divided all the 10 days' trajectories into deciles of their $\Sigma {\rm LAI}$.
They found that rainfall produced at the end of trajectories with maximum $\Sigma {\rm LAI}$ ("forest air", top decile of $\Sigma {\rm LAI}$)
is several times higher than rainfall produced by the air with minimum $\Sigma {\rm LAI}$ ("non-forest" air, bottom decile of $\Sigma {\rm LAI}$).
For each trajectory they also calculated $\Sigma {\rm ET} = \overline{\rm ET} \times t_L$ -- the total amount of moisture acquired by model-derived
evapotranspiration ET into the
air as it moves from the beginning of the trajectory to the point of observation. Then they calculated
the mean difference $\Delta \Sigma {\rm ET}$ between the forest and non-forest air trajectories and compared $\Delta \Sigma {\rm ET}$
with the mean difference $\Delta {\rm Rain}$ in daily rainfall produced at the end of the
trajectories (Fig.~\ref{spr1}). Noting that  $\Delta \Sigma {\rm ET}/ \Delta {\rm Rain} > 1$ (see Table 2 in their Supplementary Information)
\citet{spr} concluded that
forest-derived evapotranspiration "more than accounts" for the additional rain.

But $\Delta \Sigma {\rm ET}$ (kg~m$^{-2}$) and the difference in daily rainfall $\Delta {\rm Rain}$ (kg~m$^{-2}$~d$^{-1}$)
are quantities of different dimensions (i.e. units).
The value of $\Delta \Sigma {\rm ET}$ grows with increasing time length $t_L$ of the terrestrial part of the air trajectory,
while $\Delta {\rm Rain}$ is calculated for an arbitrary chosen unit of time $t_R = 1$~d (Fig.~\ref{spr1}).
Being based on measures made over arbitrary periods, the ratio $\Sigma {\rm ET}/ \Delta {\rm Rain}$ does not carry
information about atmospheric processes.

We note that the comparison of $\Delta \Sigma {\rm ET}$ with $\Delta {\rm Rain}$ provides the only quantitative basis
for \citet{spr}’s conclusion that evapotranspiration explained the key processes and patterns.
In the absence of a valid quantitative analysis correlation alone does not offer insights on causation.  The correlation between
$\Delta \Sigma {\rm LAI}$ and rainfall at the end of the trajectory found from analyses of multi-year (2001-2007) data
can be a product of synchronous changes arising from some other cause without implying any direct cause-effect relationships between the
studied parameters. \citet{spr} established that over most of the tropics the trajectories with the upper decile $\Delta \Sigma {\rm LAI}$
are associated with at least double the rainfall resulting from the trajectories with the lowest decile $\Delta \Sigma {\rm LAI}$.
This effect is shown as stippling in Fig.~2c of \citet{spr}. However, especially in those regions where rainfall is highly seasonal,
high LAI is correlated in time with high rainfall: wet season causes large-scale greening. Therefore, {\it one and the same} wind trajectory
can possess both higher LAI and produce more rainfall during the wet season than during the dry season.
This temporal concurrence between LAI and rainfall can explain why
stippling in Fig.~2c of \citet{spr} is present in even those regions (such as areas  of African deserts and
the interior of India) where the dependence between rainfall and $\Sigma {\rm LAI}$ is not statistically significant for any particular
month. Apparently in this case correlation between $\Sigma {\rm LAI}$ and rainfall reflects the fact that there is more rain
(and, hence, plant life) during the wet season than during the dry season.

\section{Post-deforestation precipitation reduction}

\begin{figure*}[h]
\centerline{
\includegraphics[width=0.6\textwidth,angle=0,clip]{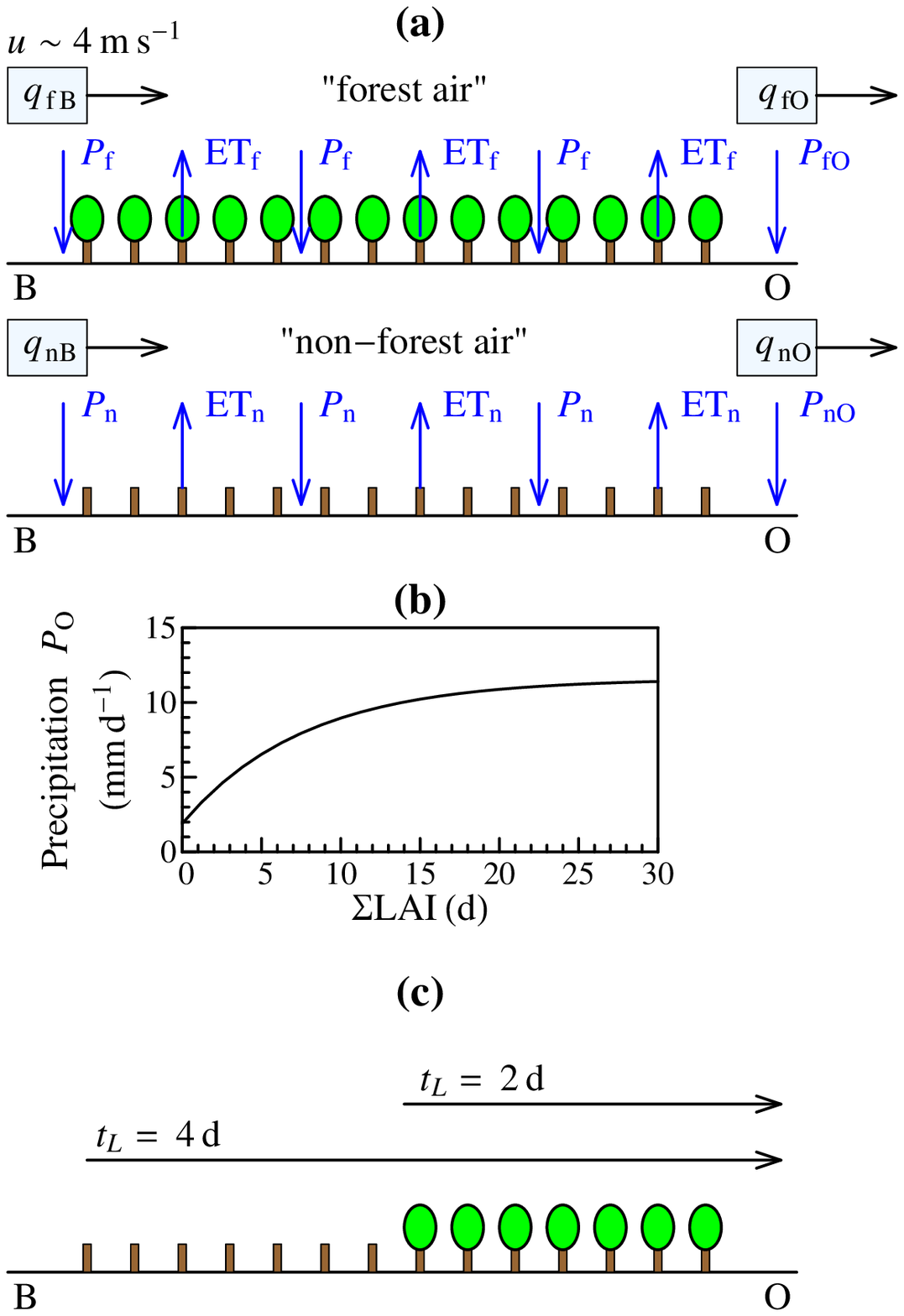}
}
\caption{\label{spr1}
(a) Simplified scheme of analysis of \citet{spr}.
Two land trajectories, one over forest and one over non-forest, are illustrated to highlight key processes.
Air with moisture content $q_{\rm fB}$ or $q_{\rm nB}$ at the beginning of the trajectory (B)
moves for $t_L$ days over the forest or non-forested land, respectively, then comes to the point of observation (O)
where precipitation rates $P_{\rm fO}$ and $P_{\rm nO}$ (kg~m$^{-2}$~d$^{-1}$) are recorded.
$\rm ET_f$ and $\rm ET_n$, $P_{\rm f}$ and $P_{\rm n}$
(kg~m$^{-2}$~d$^{-1}$) are the mean evapotranspiration and precipitation rates along the forest and non-forest trajectories,
respectively. \citet{spr} calculated $\Delta \Sigma {\rm ET = (ET_f - ET_n)} \times t_L$ and compared it with
the difference in daily precipitation at the point of observation,  $\Delta {\rm Rain} \equiv  (P_{\rm fO} - P_{\rm nO}) \times t_R$, $t_R = 1$~d.
Note that air moving at about 4 m s$^{-1}$  spends about 7 hr (much less than $t_R$) in the point of observation ($1 \times 1^{\rm o}$ cell, approx.
$100 \times 100$ km$^2$).
\newline
(b) The dependence of the observed wet season rainfall on $\Sigma \rm LAI$ established by \citet{spr} for Minas Gerais, Brazil.
\newline
(c) Different effect of deforestation on $\Sigma \rm LAI$ of air trajectories of different duration $t_L$. Removal
of half of the forest (cf. (a)) has halved $\Sigma \rm LAI$ of the 4-day trajectory arriving at point O ($t_L = 4$~d),
but left $\Sigma \rm LAI$ of the shorter 2-day trajectory unchanged.
}
\end{figure*}

\citet{spr} extended their analyses to estimate how precipitation in the Amazon
could be affected by large-scale deforestation. To make these predictions they calculated the dependence of
rainfall on 10-days' $\Sigma {\rm LAI}$  (Fig.~\ref{spr1}b).

\citet{spr} also calculated the likely LAI distribution in a deforested Amazon.
Assuming that the air circulation remains unchanged upon deforestation, they re-calculated $\Sigma {\rm LAI}$ for all
the 10-day trajectories. Then they recalculated the daily rainfall based on the established dependencies of local daily rainfall on $\Sigma {\rm LAI}$.
For example, if a trajectory had previously been characterised by $\Sigma {\rm LAI} = 10$~d but fell to $\Sigma {\rm LAI} = 3$~d because
of deforestation, then, according to the analysis of \citet{spr}, the rainfall at the end of the trajectory would decline
from about 9~mm~d$^{-1}$ to 4~mm~d$^{-1}$ (Fig.~\ref{spr1}b).
By applying this procedure to the entire Amazon basin, \citet{spr} estimated that deforestation would reduce
rainfall by 12\% during the wet season and by 21\% during the dry season.
They indicated that this reflects the loss of precipitation
recycling \citep[][Electronic Methods, http://www.nature.com/nature/journal/v489/n7415/full/nature11390.html]{spr}

However, as we have noted above, these quantitative estimates vary with respect to the chosen duration of the air trajectory $t_T$ (and of its associated terrestrial portion
$t_L$). The choice of the trajectory duration determines the spatial scale
at which deforestation "influences" rainfall.

For illustration consider the trajectory in Fig.~\ref{spr1}a.
The air covers the distance from the coast to the point of observation in $t_L = 4$~d
and the forest traversed has a mean LAI of $\rm \overline{LAI} = 3$, such that $\Sigma {\rm LAI} = 12$~d for this
particular trajectory. If half of the forest closest to the coast is converted to a desert with $\rm \overline{LAI} = 0$,
then $\Sigma {\rm LAI}$ of our {\it 4 days'} land trajectory halves to $\Sigma {\rm LAI} = 6$~d, see (\ref{sumLAI}).
The dependence established by \citet{spr} predicts a post-deforestation reduction in rainfall at the point
of observation O because of the decline in $\Sigma \rm LAI$ (Fig.~\ref{spr1}b).

Now consider air trajectories with only $t_L =2$~d. In such a case, deforestation of the coastal forest leaves
unchanged the $\Sigma {\rm LAI}$ of the {\it 2 days'} air trajectory arriving at point O.
Over the two preceding days the air will be moving over the continental interior forest and
is not affected by the coastal deforestation (Fig.~\ref{spr1}c). As $\Sigma {\rm LAI}$ has not changed, one would conclude
that there will be {\it no precipitation reduction} at point O upon the same amount of regional deforestation.
(This conclusion is invariant with respect to possible transformations of the dependence between $\Sigma {\rm LAI}$
and rainfal (Fig.~\ref{spr1}b) upon transition from longer to shorter trajectories.)
Thus, the estimated impact of regional deforestation on local precipitation depends on the duration chosen for the air trajectories. Estimates based on
such apparently arbitrary choices lack clear meaning and we are forced to conclude that the associated quantitative predictions are questionable.

To understand how forest cover influences rainfall we must understand what physical parameters determine rainfall.
Rainfall occurs when moist air ascends and cools. From simple mass balance considerations precipitation rate
$P$ can be expressed as
\beq
\label{P}
P = wq(1 -\gamma_c/\gamma_s),
\eeq
where $w$, $q$ and $\gamma_s$ are, respectively, the upward air velocity, absolute humidity and water vapor mixing
ratio at the level where condensation commences and  $\gamma_c$ is mixing ratio at a height where it stops \citep{jas13}.
In particular, if $w < 0$ (the descending air motion), any large-scale precipitation would be absent.

Ignoring turbulent admixture of moisture into the air, moisture content at the point of observation
$q_{\rm O}$ is equal to $q_{\rm O} = q_{\rm B} + \Sigma {\rm ET} - \Sigma P$ (Fig.\ref{spr1}a). It depends not only on the cumulative evapotranspiration
along the trajectory $\Sigma {\rm ET = \overline{ET}} \times t_L$, but also on the initial moisture content $q_{\rm B}$ at the beginning
{\it of the terrestrial part} of the trajectory and the cumulative terrestrial rainfall
$\Sigma P = {\overline P} \times t_L$ that depletes the air moisture content.
We note that this rainfall was
neglected by \citet{spr} in their analyses, as were the evaporation and precipitation processes associated with the oceanic part of each trajectory.

\citet{spr} noted that moisture content diminishes less in the forest air than in the non-forest air from the beginning of the
land part of the trajectory: $(q_{\rm fO} - q_{\rm fB}) > (q_{\rm nO} - q_{\rm nB})$.
However,
three additional requirements must be met in order to demonstrate that the additional rain is explained wholly by the air's
higher moisture content due to forest evapotranspiration.  First, the intensity of convection (described by vertical air velocity
and the completeness of condensation in the atmospheric column) must be equal at the point of observation for the forest and non-forest air.
Second, $q_{\rm O}$ must be higher in the forest versus non-forest air ($q_{\rm fO} > q_{\rm nO}$).
Third, $\Sigma {\rm ET}$ (and not $q_{\rm B}$ or $\Sigma P$) must solely determine this difference.

Such an analysis would certainly be difficult. While TRMM rainfall, wind directions and LAI are empirically observable variables,
evapotranspiration $\rm ET$ is model-derived. Moisture flux from the tropical vegetation cover depends
on a large number of poorly known biotic processes and thus its estimate presents a great challenge \citep{fisher09,restrepo13}.
Other complications reflect the shortcomings of models in representing atmospheric moisture transport.
Modelling studies may disagree with some lines of observational evidence concerning the predominant
direction of regional moisture transport. For example, for the Congo rainforest model reconstructions suggested that moisture
comes to the region predominantly from the East \citep{ent10}, while the available evidence on atmospheric pressure gradients
and isotope data appear to testify that the rainforest receives its moisture from the West (the Atlantic ocean)
and is rather a source of moisture for Eastern Africa \citep{nicholson00,williams12}.
Note also the mismatch between the modelled atmospheric moisture convergence and actual observed runoff seen in
the studies of the Amazon water budget \citep[e.g.,][Table~4]{ma06}.  Here it is pertinent that the atmospheric moisture convergence for the Amazon basin
derived from model reanalyses (e.g., NCEP/NCAR) is about half the figure estimated from runoff observations. This implies that modeled parameters are
distorted or incomplete and should be treated with caution. If the source models used to estimate evapotranspiration underestimate actual moisture
convergence but still reproduce rainfall satisfactorily, this would imply that evapotranspiration is overestimated.

\section{What determines rainfall?}

\subsection{Comparing rainfall and pressure for winds of different directions}

Noting that any of the parameters in (\ref{P}) might account for the rainfall differences between the forest and non-forest air,
we can formulate an alternative, and fully consistent, explanation for the empirical patterns established by \citet{spr}.
Recently we have proposed that natural forest cover can cause low atmospheric pressure.
The mechanism derives from evaporation and condensation and resulting gradients in atmospheric moisture -- in brief the areas with
the highest evaporation drive upwelling and condensation, that induces low pressure and draws in most air from elsewhere
leading to a net atmospheric moisture inflow to the continent from the ocean (see \citep{mg07} for the basic ideas and \citep{m13} and
references therein for a fuller account of the physical
principles behind it). Evidence for these mechanisms has already come from a number of studies showing how rainfall over non-forested
areas tends to decrease exponentially with increasing distance from the ocean, while it stays more constant over forests
\citep{mg07,doug09,mgl09,mgl13,poveda13}.

Based on the above theory the air coming from a forest-covered region
produces more rain because it is often associated with low pressure systems (ascending air motion $w >0$, high rainfall), while the air
that arrives from non-forest regions is more often associated with high(er) pressure systems (descending air motion $w < 0$, low rainfall).
To investigate these propositions we examined the relationships among rainfall, atmospheric pressure and wind direction in three regions
in tropical South America (Fig.~\ref{fig2}). Regions A (15-20$^{\rm o}$~S, 45-40$^{\rm o}$~W) and B (5-10$^{\rm o}$~S, 40-45$^{\rm o}$~W)
correspond to the two areas where, according to Fig.~2c of \citet{spr}, the relationship between rainfall and
$\Sigma {\rm LAI}$ in South America is the strongest: it is present during more than 8 months of a year. Region A and Region C belong to two of the four
tropical areas analyzed in detail by \citet{spr}.
Region C is the core of the Amazon forest where, according to Fig.~2c of \citet{spr}, there is little if any dependence
between $\Sigma {\rm LAI}$ and rainfall.

In each region we investigated 14 meteorological stations with pressure, wind direction and precipitation records
provided by the Brazilian Meteorological Institute\footnote{http://www.inmet.gov.br/portal/index.php?r=bdmep/bdmep}. Wind direction and air pressure
are measured at 00, 12 and 18 hours daily and precipitation data are provided on a daily basis. For each station, we took all the available data up to
31 December 2012, the earliest date being 2 January 1961 (see Table 1 in Supplementary Information\footnote{\url{http://www.bioticregulation.ru/common/pdf/spr/sprn-sup.pdf}} for specific data for each station.)
Total number of day/hour combinations is 602,083 for region A, 380,091 for region B and 340,484 for region C.
The Amazon forest is located to the West and North-West of the non-forest regions A and B (Fig.~\ref{fig2}). We divided all wind direction data into four
categories, having combined West and North-West (WNW), East and South-East (ESE), North and North-East (NNE), South and South-West (SSW) winds.

For each of the three daily observation hours during the dry and wet season we
compared mean rainfall and pressure during periods of West-North-West versus East-South-East winds (rows "WNW-ESE" in Fig.~\ref{fig3}).
We thus obtained 84 pressure difference values (the green bars) and 84 rainfall difference values (the brown bars) for each region
(six pressure and six rainfall difference values for each of the 14 stations in each region: 00, 12, 18 hours during the wet season (the first
triplet of bars) and 00, 12, 18 hours during the dry season (the second triplet of bars).)
For each station, we defined the wet season as the six months with the maximum rainfall and the dry season as the rest of the year.
The wet season was from October through March and from November through April for all stations in regions A and B, respectively. In region C
when averaged over all stations the wet season is from December through May, although there are differences between some stations.
In Fig.~\ref{fig2}b calculations are made for the region mean, while in Figs.~\ref{fig3} and~\ref{fig4} for each station in region C its own wet season is
considered (see Table 1 in Supplementary Information for details).

We applied the Student $t$-test to test the null hypothesis of there being zero difference in the mean WNW and ESE pressure and rainfall.
We found that those days when the wind arrives to regions A and B from West or North-West have consistently higher rainfall
than the days with the wind blowing from the East or South-East (i.e. from a direction opposite to the Amazon forest).
The WNW$-$ESE rainfall differences $\Delta P$ are predominantly positive. In region A, out of 84 values only two $\Delta P$ are negative
and none of them are statistically significant non-zero at 0.01 probability level; 71 values are positive and statistically significant (Table~\ref{t1}).
Likewise in region B, only two out of 84 values are negative and none of them statistically significant, 82 are positive and
and 68 of them are statistically significant.
These findings are consistent with the air coming from the forest bringing more rain
to regions A and B than the air coming from the ocean (East and South-East direction). However, it should be noted that in both regions WNW winds are less frequent than ESE winds
(Fig.~\ref{fig2}), such that days with WNW winds make
a markedly smaller contribution to the annual rainfall than the ESE winds even though they are typically rainier when they occur (see Table 2 in Supplementary Information
for detailed accounts for each station).

\begin{table}[t]
\caption{Rainfall ($\Delta P$) and pressure ($\Delta p$) differences between West and North-West (WNW) versus East and South-East (ESE) winds
and between South and South-West (SSW) versus North and North-East (NNE) winds in Regions A, B and C (Fig.~\ref{fig3}). Numbers represent statistically
significant cases at 0.01 significance level. The maximum possible number is 84 (four values for each of the 14 stations in each region).
}\label{t1}
\begin{center}
\begin{tabular}{llcrrrrrr}
\hline
Region & Wind pair & $\Delta P > 0$ & $\Delta P \le 0$ & $\Delta p < 0$ & $\Delta p \ge 0$ & $\Delta P \Delta p < 0$ & $\Delta P \Delta p \ge 0$ \\
\hline
A& WNW$-$ESE & 71 & 0 & 80 & 1 & 68 & 0 \\
A& SSW$-$NNE & 17 & 16 & 36 & 26 & 20 & 4 \\
B& WNW$-$SSE & 68 & 0 & 54 & 5 & 48 & 3 \\
B& SSW$-$NNE & 24 & 7 & 8 & 54 & 9 & 13 \\
C& WNW$-$SSE & 22 & 4 & 24 & 16 & 3 & 11 \\
C& SSW$-$NNE & 26 & 1 & 3 & 58 & 0 & 20 \\
\hline
\end{tabular}
\end{center}
\end{table}

Comparison of the other pair of wind directions, South-South-West versus North-North-East, which represent either non-forested territories for region A
or non-forested territories (SSW) versus the ocean (NNE) for region B, does not reveal a considerable difference in rainfall rates.
In region A only 33 $\Delta P$ values are statistically significant, of which 17 are positive and 16 are negative. In region B,
these numbers are, respectively, 31, 24 and 7, showing slightly higher rates associated with continental SSW winds. However, here
as well the SSW winds are infrequent and their impact on annual precipitation is small.

Second, in agreement with our proposition the days with WNW ("forest") winds in regions A and B are characterized by a consistently lower pressure
than the days with ESE ("non-forest") winds. In region A, 83 $\Delta p$ values are negative (80 significant) and only one is positive
and significant. In region B, 72 $\Delta p$ values are negative (54 significant) and only 12 are positive (five significant).
This analysis provide support for the statement that the influence of different wind directions on rainfall cannot be reduced
to the moisture content alone, but involves changes in atmospheric pressure and potentially other parameters.
It also shows that given there is consistent difference in rainfall associated with winds blowing from different directions,
any parameter correlated with these spatial directions will be consistently correlated with rainfall.
In particular, as the air arriving to region A with the North-West winds from the Amazon forest is associated with higher rainfall than
the air brought by the East winds coming from the ocean, any parameter that is consistently differentiating ocean and forest (be that LAI or,
for example, surface roughness) will be similarly correlated with higher or lower rainfall. Apparently, such a correlation does not {\it per se}
presume a cause-effect relationship.

The situation in region C is somewhat different. First, there is less difference in the mean daily rainfall
derassociated with the different wind directions:
only 26 and 27 $\Delta P$ values are statistically significant in WNW$-$ESE and SSW$-$ENE comparisons, respectively (Table \ref{t1}),
with western directions (WNW and SSW) tending to have higher rainfall. Second, the WNW and ESE winds do not differ much in their
pressure either (out of 40 significant $\Delta p$ values 24 are positive and 16 are negative). Meanwhile the North-North-East direction (the
predominant direction of oceanic moisture transport inland) is associated with a significantly higher atmospheric pressure than the opposite South-South-West
direction (58 out of 61 statistically significant $\Delta p$ values are positive).

In region C the majority of cases in which both $\Delta P$ and $\Delta p$ are non-zero, higher rainfall is associated with higher atmospheric pressure
($\Delta P \Delta p >0$), compare the last two columns in Table~\ref{t1}.
In regions A and B, out of respectively 92 and 73 cases where $\Delta P$ and $\Delta p$ are {\it both} significant
(WNW$-$ESE and SSW$-$NNE comparisons combined), they have different signs in 88 (A) and 57 (B) cases.
The dominant pattern in these two non-forest regions is thus lower pressure at higher rainfall.
In contrast, in forested region C only 3 cases out of 31 significant pairs have different signs for $\Delta P$ and $\Delta p$.

\subsection{Comparing pressure in rainy versus rainless days}

We compared atmospheric pressure between rainy and rainless (zero recorded precipitation) days in the three regions (Fig.~\ref{fig4}). During rainy days the atmospheric
pressure in Region A averages
1-2 hPa lower than during rainless days. In the dry season the effect is roughly twice what is observed over the wet season. In contrast,
in the forest region (Region C), pressure is 0.5-1 hPa {\it higher} during rainy days than during rainless days. This effect is, contrary to the results from Region A,
less pronounced in the dry season.
Region B, located closer to the Amazon forest than region A, displays an intermediate pattern between A and C: during the wet season the pressure
is higher, and during dry season it is lower, in rainy versus rainless days.

The diurnal cycle of precipitation over tropical land peaks in the early afternoon, which is approximately
concurrent with the diurnal minimum of surface pressure \citep[e.g.,][]{silva87,lin00,yang06}.
Thus, surface pressure
measured during precipitation events should be on average lower than the long-term daily mean. However,
as our analysis has revealed (Fig.~\ref{fig4}c), in the forest region C the mean surface pressure during rainy days is consistently
{\it higher} at both 00, 12 and 18 hours local time than it is at the corresponding hours during rainless days. This means
that the daily barometric minimum during rainy days in the forest region C might be shallower (or the barometric maximum steeper)
than it is during rainless days. The opposite pattern should take place
in region A year-round and in region B during the dry season.
To our knowledge, these differences in the rainfall/pressure relationship have not been previously described.
As we discuss below, they indicate different processes of precipitation formation in the forested
and non-forested regions.

Our analysis of radiosonde data that are provided by the University of Wyoming\footnote{http://weather.uwyo.edu/upperair/sounding.html} for several stations from regions A, B and C revealed that
the difference in column water vapor (CWV) between rainy and rainless days is, as might be expected, larger during the dry
than during the wet season (Table~\ref{t2}). For example, in the wet season in Manaus (station 6 in region C) mean CWV is
2 mm larger on rainy than on rainless days (58 mm vs 56 mm), while the corresponding difference in the dry season is 4 mm
(53 mm vs 49 mm). Note that in the wet season
in region C rainy days are about twice as frequent as rainless days, while in the dry season the opposite is true with dry days being twice as frequent as rainy days.
In Brasilia (station 2 in region A) wet season CWV during rainy days is 7.5 mm larger (35 mm vs 27.5 mm), in the dry season
it is 11.5 mm larger (30.8 mm vs 19.3 mm) than during the dry season (Table~\ref{t2}). The proportion of rainy to rainless days
is  1:6 in the dry season and 3:2 in the wet season (Table~\ref{t2}).

\begin{table}[t]
\caption{Rainfall $P$, pressure $p$ and column water vapor CWV differences between rainy ($+$) and rainless ($-$) days
in several stations from regions A, B, C with available radiosonde data (see Table 3 in the Supplementary Information for details). {\it t} -- time
of radiosonde measurement of surface pressure, $N$ -- number of observations, $\overline{P}$ -- mean
rainfall (mm~d$^{-1}$), $\overline{p}$ -- mean pressure $\pm$ standard deviation,
$l^+$ -- proportion of rainy days, $l^-$ -- proportion of rainless days, $p^+$ ($p^-$) -- mean pressure at time $t$ during
rainy (rainless) days (hPa), $\Delta p \equiv p^+-p^-$ (hPa), CWV$^+$ (CWV$^-$) -- column water vapor
during rainy (rainless) days (mm), $\Delta {\rm CWV} \equiv {\rm CWV}^+ - {\rm CWV}^-$ (mm).
All $\Delta p$ and $\Delta {\rm CWV}$ values differ significantly from zero at 0.01 probability level (Student $t$-test)
except for the two $\Delta p$ values for the dry season in station C11.
}\label{t2}
\begin{center}
\begin{tabular}{lllrrrrrrrrrrr}
\hline
\scriptsize Station &\scriptsize Season/$t$ &\scriptsize $N$ &\scriptsize $\overline{P}$ &\scriptsize $\overline{p}$ &\scriptsize $l^+$ &\scriptsize $l^-$ &\scriptsize  $p^+$ &\scriptsize $p^-$ &\scriptsize$\Delta p$ &\scriptsize CWV$^+$ &\scriptsize CWV$^-$ &\scriptsize $\Delta {\rm CWV}$ \\
\hline
\scriptsize A2  Brasilia         &\scriptsize  WET/00  &\scriptsize 2755    &\scriptsize   6.4    &\scriptsize    895.4$\pm$1.8  &\scriptsize   0.41   &\scriptsize    0.59  &\scriptsize     895.29  &\scriptsize   895.66  &\scriptsize   -0.38  &\scriptsize    37.8  &\scriptsize     29.4  &\scriptsize     8.39  \\
\scriptsize A2  Brasilia         &\scriptsize  WET/12  &\scriptsize 5591    &\scriptsize   6.7    &\scriptsize    897.4$\pm$1.8  &\scriptsize   0.41   &\scriptsize    0.59  &\scriptsize     897.09  &\scriptsize   897.81  &\scriptsize   -0.73  &\scriptsize    35.1  &\scriptsize     27.5  &\scriptsize     7.57  \\
\scriptsize A2  Brasilia         &\scriptsize  DRY/00  &\scriptsize 2955    &\scriptsize   1.2    &\scriptsize    898.6$\pm$2.0  &\scriptsize   0.87   &\scriptsize    0.13  &\scriptsize     897.15  &\scriptsize   898.81  &\scriptsize   -1.7   &\scriptsize    33.9  &\scriptsize     20.7  &\scriptsize     13.2  \\
\scriptsize A2  Brasilia         &\scriptsize  DRY/12  &\scriptsize 5422    &\scriptsize   1.3    &\scriptsize    900.2$\pm$2.0  &\scriptsize   0.86   &\scriptsize    0.14  &\scriptsize     898.89  &\scriptsize   900.39  &\scriptsize   -1.5   &\scriptsize    30.8  &\scriptsize     19.3  &\scriptsize     11.5  \\
\scriptsize B4 Floriano          &\scriptsize  WET/12   &\scriptsize 1715   &\scriptsize   5.     &\scriptsize    998.9$\pm$1.3  &\scriptsize   0.57   &\scriptsize    0.43  &\scriptsize     999.03  &\scriptsize   998.81  &\scriptsize   0.22   &\scriptsize    47.5  &\scriptsize     40.4  &\scriptsize     7.11  \\
\scriptsize B4 Floriano          &\scriptsize  DRY/12   &\scriptsize 1588   &\scriptsize   0.82   &\scriptsize    1000.7$\pm$1.9  &\scriptsize  0.90   &\scriptsize    0.10  &\scriptsize     999.54  &\scriptsize   1000.8  &\scriptsize   -1.3   &\scriptsize    45.6  &\scriptsize     32.6  &\scriptsize     12.9  \\
\scriptsize B14 Petrolina        &\scriptsize  WET/12   &\scriptsize 1889   &\scriptsize   2.1    &\scriptsize    971.6$\pm$1.5  &\scriptsize   0.82   &\scriptsize    0.18  &\scriptsize     971.16  &\scriptsize   971.63  &\scriptsize   -0.47  &\scriptsize    41.7  &\scriptsize     32.5  &\scriptsize     9.2   \\
\scriptsize B14 Petrolina        &\scriptsize  DRY/12   &\scriptsize 1926   &\scriptsize   0.27   &\scriptsize    974.8$\pm$2.0  &\scriptsize   0.93   &\scriptsize    0.07  &\scriptsize     974.38  &\scriptsize   974.85  &\scriptsize   -0.48  &\scriptsize    34.2  &\scriptsize     25.8  &\scriptsize     8.39  \\
\scriptsize C6  Manaus           &\scriptsize  WET/00  &\scriptsize 1962    &\scriptsize   9.4    &\scriptsize    1000.1$\pm$1.6 &\scriptsize   0.35   &\scriptsize    0.65  &\scriptsize     1000.3  &\scriptsize   999.76  &\scriptsize   0.55   &\scriptsize    58.2  &\scriptsize     56.5  &\scriptsize     1.74  \\
\scriptsize C6  Manaus           &\scriptsize  WET/12  &\scriptsize 4345    &\scriptsize   9.2    &\scriptsize    1002.5$\pm$1.4 &\scriptsize   0.34   &\scriptsize    0.66  &\scriptsize     1002.6  &\scriptsize   1002.2  &\scriptsize   0.43   &\scriptsize    54.8  &\scriptsize     52.4  &\scriptsize     2.4   \\
\scriptsize C6  Manaus           &\scriptsize  DRY/00  &\scriptsize 2311    &\scriptsize   3.4    &\scriptsize    1000.7$\pm$2.1 &\scriptsize   0.69   &\scriptsize    0.31  &\scriptsize     1001.0   &\scriptsize   1000.5  &\scriptsize   0.54   &\scriptsize    52.8  &\scriptsize     48.6  &\scriptsize     4.15  \\
\scriptsize C6  Manaus           &\scriptsize  DRY/12  &\scriptsize 4832    &\scriptsize   3.3    &\scriptsize    1003.2$\pm$1.8 &\scriptsize   0.68   &\scriptsize    0.32  &\scriptsize     1003.4  &\scriptsize   1003.1  &\scriptsize   0.34   &\scriptsize    49.8  &\scriptsize     45.4  &\scriptsize     4.48  \\
\scriptsize C11 C. do Sul        &\scriptsize  WET/00  &\scriptsize 304     &\scriptsize   8.5    &\scriptsize    986.0$\pm$2.1  &\scriptsize   0.38   &\scriptsize    0.62  &\scriptsize     986.31  &\scriptsize   985.56  &\scriptsize   0.75   &\scriptsize    55.9  &\scriptsize     53.3  &\scriptsize     2.55  \\
\scriptsize C11 C. do Sul        &\scriptsize  WET/12  &\scriptsize 1073    &\scriptsize   7.5    &\scriptsize    988.5$\pm$1.8  &\scriptsize   0.38   &\scriptsize    0.62  &\scriptsize     988.72  &\scriptsize   988.17  &\scriptsize   0.55   &\scriptsize    54.3  &\scriptsize     51.4  &\scriptsize     2.85  \\
\scriptsize C11 C. do Sul        &\scriptsize  DRY/00  &\scriptsize 275     &\scriptsize   3.7    &\scriptsize    988.1$\pm$2.4  &\scriptsize   0.64   &\scriptsize    0.36  &\scriptsize     987.81  &\scriptsize   988.23  &\scriptsize   -0.42 &\scriptsize    53.1  &\scriptsize     47.9  &\scriptsize     5.17  \\
\scriptsize C11 C. do Sul        &\scriptsize  DRY/12  &\scriptsize 1176    &\scriptsize   3.8    &\scriptsize    990.7$\pm$2.2  &\scriptsize   0.64   &\scriptsize    0.36  &\scriptsize     990.56  &\scriptsize   990.72  &\scriptsize   -0.16 &\scriptsize    49.8  &\scriptsize     44.1  &\scriptsize     5.69  \\
\hline
\end{tabular}
\end{center}
\end{table}

\citet{holloway10} using the 1-min resolution long-term data record for an equatorial island station showed that when CWV is larger than 30 mm the probability of subsequent rainfall
rises sharply with increasing CWV. At CWV between 20 and 30 mm the
rainfall probability does not appear to relate to variation in CWV (Fig.~\ref{fig5}).
The feedback between atmospheric water vapor and convection processes have been investigated
and discussed \citep[e.g.,][]{bretherton04,holloway09,holloway10}. On the one hand, the local rise of moist saturated air that undergoes condensation
can enrich a previously drier upper atmosphere with water vapor leading to growing CWV. On the other hand,
a high CWV has itself been proposed as a trigger of larger-scale convection \citep[e.g.,][]{mapes93,grabowski04,sharkov12}.
In our recent work we proposed that a moist atmosphere is unstable owing to the non-equilibrium pressure gradients that arise in the
ascending air because of water vapor condensation \citep{mg07,g12,m13,jas13}. Here we offer our ideas how the observed contrasting
rainfall/pressure patterns in forested and non-forested regions can be explained from such a perspective.

Precipitation depletes atmospheric moisture, while evaporation replenishes this loss.
Since condensation depends on the vertical velocity of the ascending saturated air it can theoretically occur at an arbitrarily high rate, while
the rate of surface evaporation depends on solar radiation flux and is limited.
Consider an area with a length scale exceeding $10^3$~km (and neglect for now any exchange of air with the surroundings).
If the air is moist and rises high in one place and descends in another within the considered area (a
deep convection event), this will lead to the reduction of
mean pressure in the area owing to the removal of precipitated moisture in the ascending branch from the atmospheric column. This process will end once
the atmosphere has become sufficiently dry. The next major condensation/precipitation event will not occur until water vapor has accumulated in the
atmosphere beyond the critical limit. Such accumulation occurs via evaporation, which is a relatively slow process. This gradual build-up of moisture
will be accompanied by gradual rise of surface pressure in the area reflecting the growing mass of the atmospheric column.
Thus, most intense precipitation in our area will occur when the CWV and, consequently, the {\it mean surface pressure} in the area are at their highest.
Rainless days will occur when CWV and pressure are diminished by precipitation, and will be characterized by lower pressure.

Besides these {\it temporal} changes of mean pressure associated with build-up and removal of atmospheric water vapor,
there are also {\it spatial} pressure differences between the region of ascent (air convergence, lower pressure)
and descent (air divergence, higher pressure). In a very humid region like the rainforest
there are small rains even in the descending branch of a larger-scale circulation driven by small-scale vortices (e.g., those leading to the
formation of convective rolls (cloud streets) that have a space scale of the order of a few kilometers \citep{nair03,ramos11}). Condensation
in the descending branch makes the pressure difference between the ascending and descending branches less pronounced and the associated
winds milder. Indeed, some moisture evaporated in the descending branch precipitates locally rather than adds to the pressure
difference between the ascending and the descending branch. This prevents formation of strong winds (that would appear when the pressure
gradient is large) and makes precipitation events spatially and temporarily uniform \citep{millan11}.
In the dry environments precipitation events are rare and more extreme. Because of low humidity shallow convection in the descending branch is absent,
while the pressure difference between ascent and descent is higher.

Comparing atmospheric pressure during rainy and rainless days in such different regions we should expect to find that
in the dry region the pronounced {\it spatial} differences
between the ascending and descending convection branches will dominate: rainfall will be recorded on the station if the station finds itself
in the low pressure convergence area. The relatively infrequent rainy days will have lower pressure than the more frequent rainless days.
In the humid region the
temporal pressure changes should dominate: pressure should be lowest during the few rainless days following the dehumidification of the atmosphere by a major
precipitation event.
The majority of days before the next major precipitation event will be rainy and see higher pressure.
The limited quantitative evidence available agrees with such a pattern.
\citet{holloway10} showed that approximately 36 hours prior to
a major precipitation event
($> 0.97$~mm~hr$^{-1}$) CWV accumulates on a synoptic scale ($\sim 10^3$~km) at a rate of about 2.5 mm~d$^{-1}$. This coincides
in magnitude with the rate of evaporation in the Amazon \citep{ma05}. A build-up of 5 mm of CWV will increase surface pressure by 0.5 hPa
(1 mm $\approx$ 0.1 hPa), which is a characteristic difference between rainy and rainless days observed in Region C (Fig.~\ref{fig3}). The
observed difference in CWV between rainy and rainless days during the wet season in region C is about 2 mm (Table 2), which
explains about half of the observed pressure difference. Fine-scale measurements
would be valuable, as in for example \citep{holloway10}.

\subsection{The time scale of rainfall/pressure coupling}

The fact that higher pressure during rainy days is an effect that is pronounced on a short time scale
is further supported by the following analysis. Atmospheric pressure $p$ and rainfall $P$ on each station undergo significant
seasonal changes (Fig.~\ref{fig6}). Their correlation in time can exert influence on patterns shown in Fig.~\ref{fig3}. For example,
if pressure during the wet season is on average higher than during the dry season, then, when averaged over
all observations, rainy days will likely have higher pressure just because they are more numerous in the wet season
when the pressure is high. Conversely, if the wet season pressure is lower than the dry season pressure, then
rainless days will have on average higher pressure.

To gain insight into these effects, for each station we took pressure values corresponding to 12 hr (00 and 18 hr give similar results)
and smoothed them by using moving averages based on periods of $m =$ 3, 5, 7, 9, 11, 13, 17, 19, 21, 25, 31, 35, 41, 61, 91, 121, 183 and
365 days over the uninterrupted spells of observations of length greater than or equal to $m$.
For example, for $m=7$ we calculated the mean pressure for the week (seven days) centered at each given day:
e.g., the smoothed value for 10 April will be the mean of pressure values measured at 12 hr on 7, 8, 9, 10, 11, 12 and 13 April.
For every $i$-th day and a given $m$, we calculated difference $\Delta p_i \equiv p_i-\overline{p}_{i}(m)$ between the actual value of pressure
measured during this day and the smoothed value. We considered $\Delta p_i$ values for only those rainy days
that were accompanied by at least one rainless day
(and for only those rainless days that were accompanied by at least one rainy day) in the spell of $m$ days centered at the considered day.
Hence, the  difference $\Delta p_i$ for 10 April was considered if it was a rainy (rainless) day and at least one day out of 7, 8, 9, 11, 12 or
13 April was rainless (rainy). We then compared all the obtained differences with zero separately for rainy and rainless days.

The analysis showed that in the forest region C the effect of
higher pressure during rainy days was present already at $m=3$~d (i.e. moving averages over 3 days). E.g. in Manaus
at $m=3$ we compared pressure of $N^+ = 4359$ rainy days with the corresponding smoothed means and found that in $n^{+}= 2450$ (56\%) cases
$\Delta p_i > 0$: the rainy days have higher pressure than the 3-days' mean including at least one (at most $m-1$) rainless day.
This result deviates from the null hypothesis mean of $N^+/2$ by almost six standard deviations $\sigma$.
Indeed, assuming equally probable positive and negative $\Delta p_i$ values
the value of $\sigma$ would be $\sigma = \sqrt{N^+/2} = 47$: $n^{+} = N^+/2 + 5.8 \sqrt{N^+/2}$. We define $s^+(m)\equiv (n^+ - N^+/2)/\sqrt{N^+/2}$ as a measure of
significance of the effect.
We also compared pressure measured on $N^- = 4225$ rainless days with their smoothed means
and found that in $n^{-} = 2388$ (56\%) cases rainless days had lower pressure than the smoothed mean.
The effect is also significant: $s^-(m)\equiv (n^- - N^-/2)/\sqrt{N^-/2}$, for Manaus $s^-(7) = 6.0$.
These two sets of comparisons are of course not independent but neither they are identical:
the rainless days at large $m$ are more representative of the dry season. For example, one dry season month with 28 rainless
and two rainy days will contribute two values to $N^+$ and 28 values to $N^-$.

We can see that with increasing smoothing interval $m$ the significance remains the same in almost all stations in region C (Fig.~\ref{fig7}).
At very large $m=365$~d (1 year) there is a tendency for decreasing significance $s^+$ (Fig.~\ref{fig7}) and $s^-$ (data not shown).

The situation is different in regions A and B. For example, in Brasilia (station 2 in region A) at small $m = 3$ there is an insignificant tendency
for rainy days to have higher pressure and rainless days to have lower pressure. With increasing $m$ the situation changes and already at $m=17$~d the
rainy days have significantly lower pressure and rainless days have significantly higher pressure than their smoothed mean: $s^+$ and $s^-$
are both less than $-3$ (Fig.~\ref{fig7}). With increasing $m$ the significance grows and reaches maximum at $m=365$~d. This apparently
illustrates the inclusion of the seasonal correlation between the high pressure during rainless season (Fig.~\ref{fig6}).

In Fig.~\ref{fig7} the dependence of $s$ on $m$ is shown for all stations. One can see that indeed the effect of
higher pressure during rainy days is best pronounced in forest stations at almost all values of $m$ and in non-forest stations
at small values of $m$. Note that small $m$ are by construction correlated with relatively wet periods, because at least one day
in every $m$ averaged days must be rainy.
With growing time scale $m$ in non-forest regions rainy days tend on average to have lower pressure.
These patterns certainly need more study. Here however we want to emphasize the tight interplay between rainfall
and pressure on a variety of scales that determines regional atmospheric dynamics.

\section{Discussion}

We have re-analyzed the results of \citet{spr} who found that in the tropics the air that had passed over the forest brings more
rain to the non-forested regions than the air that had come from drier land regions or the ocean.
They assumed that circulation patterns are not affected by forest cover, meaning that the influence of forest cover change is
manifested as changes in atmospheric moisture content via evapotranspiration (moisture recycling).
In Sections 2 and 3 we discussed methodological problems with
the analysis and its specific quantitative conclusions regarding predicted changes in rainfall arising from forest cover loss in the Amazon region.

Studies of the potential climatic impact of deforestation have long recognized that
a change in vegetation cover affects multiple atmospheric parameters: it affects the surface energy balance through changing net surface albedo,
and surface heat capacity, affecting surface temperature and the partitioning of absorbed energy between latent, sensible, and ground heat fluxes.
Surface roughness is also reduced dramatically upon conversion of tall forests to pastures. The impact of these
parameters on atmospheric circulation has been investigated in modelling studies \citep[e.g.,][]{shukla90,sa91,polcher94}, see also
a summary by \citet[][Table V]{mcguffie01}.
In parallel, studies of the relationship between the intensity of convection and atmospheric moisture content have suggested
from different perspectives that high moisture content in the atmospheric column can trigger large scale ascending motion and condensation
\citep{mapes93,grabowski04,bretherton04,holloway09,holloway10,degu11,sharkov12}. Finally, we ourselves have proposed that forest cover can, via moisture evaporation and condensation, generate the large-scale pressure gradients,
that create the ocean-to-land moisture flux convergence that supplies rainfall \citep{mg07,g12,m13}.

\begin{figure*}[h]
\centerline{
\includegraphics[width=0.99\textwidth,angle=0,clip]{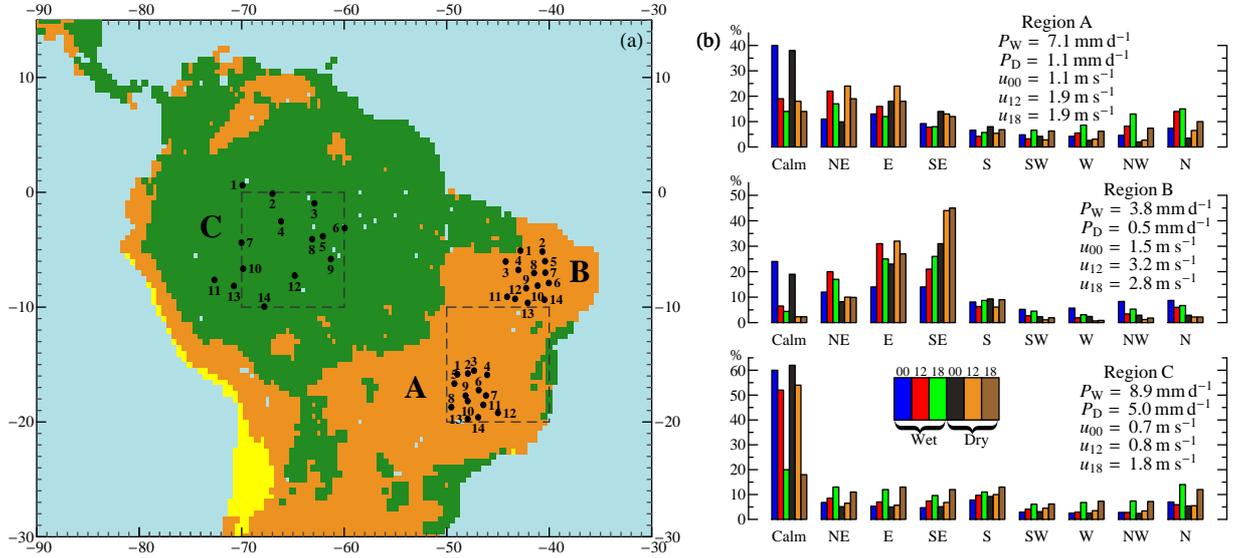}
}
\caption{\label{fig2}
(a) Stations studied in Regions A, B and C.
Station names with their WMO codes are as follows.
A:
1, Piren\'{o}polis, 83376;
2, Brasilia, 83377;
3, Formosa, 83379;
4, Arinos, 83384;
5, Goi\^{a}nia, 83423;
6, Paracatu, 83479;
7, Jo\~{a}o Pinheiro, 83481;
8, Capin\'{o}polis, 83514;
9, Ipameri, 83522;
10, Catal\~{a}o, 83526;
11, Patos de Minas, 83531;
12, Pomp\'{e}u, 83570;
13, Uberaba, 83577;
14, Arax\'{a}, 83579;
B:
1, Teresina, 82578;
2, Crate\'{u}s, 82583;
3, Colinas, 82676;
4, Floriano, 82678;
5, Tau\'{a}, 82683;
6, Ouricuri, 82753;
7, Campos Sales, 82777;
8, Picos, 82780;
9, S\~{a}o Jo\~{a}o do Piaui, 82879;
10, Paulistana, 82882;
11, Bom Jesus do Piaui, 82975;
12, Caracol, 82976;
13, Remanso, 82979;
14, Petrolina, 82983;
C:
1, Iauaret\^{e}, 82067;
2, S. G. da Cachoeira, 82106;
3, Barcelos, 82113;
4, Fonte Boa, 82212;
5, Codaj\'{a}s, 82326;
6, Manaus, 82331;
7, Benjamin Constant, 82410;
8, Coari, 82425;
9, Manicor\'{e}, 82533;
10, Eirunep\'{e}, 82610;
11, Cruzeiro do Sul, 82704;
12, L\'{a}brea, 82723;
13, Tarauac\'{a}, 82807;
14, Rio Branco, 82915.
Dashed
squares indicate the two South American regions (Amazon and Minas Gerais) discussed by \citet{spr}.
Vegetation cover (forest, green; non-forest vegetation, brown; unvegetated land, yellow)
is shown following \citet{friedl10} (see \citet[][Fig.~1]{mgl13} for details).
\newline
(b) Mean wind frequencies in the two regions in the dry and wet season
at different times of the day. Heights of bars of the same color sum up to 100\%.
Also shown are mean wet season ($P_{\rm W}$) and dry season ($P_{\rm D}$) precipitation
and mean annual wind velocity $u$ measured at 00, 12 and 18 hr.
See Table 2 in the Supplementary Information for
data for each station.
}
\end{figure*}

\begin{figure*}[h]
\centerline{
\includegraphics[width=0.99\textwidth,angle=0,clip]{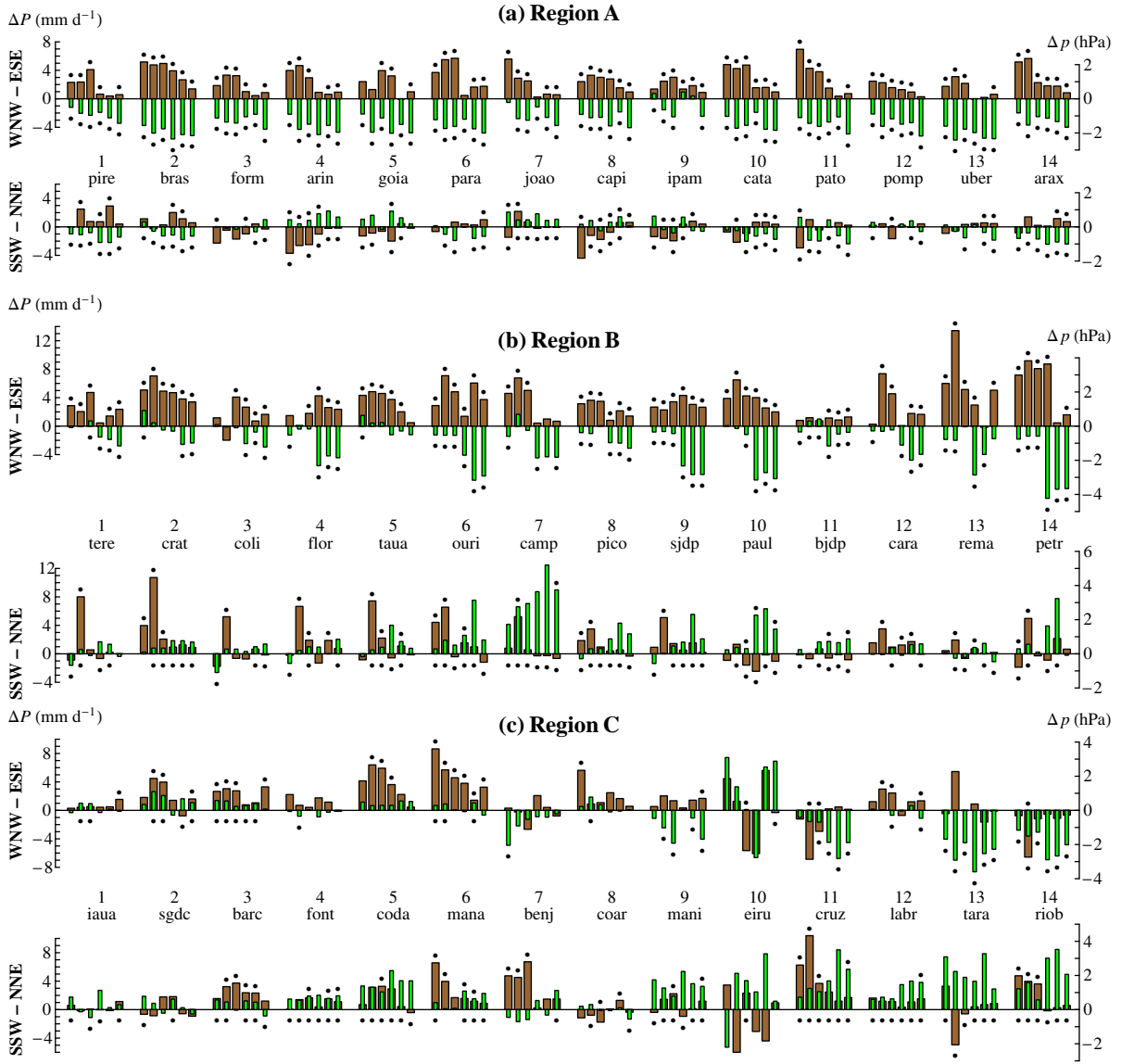}
}
\caption{\label{fig3}
Mean difference in daily rainfall ($\Delta P$, thick brown bars) and surface pressure ($\Delta p$, thin green bars)
between days with winds of different directions blowing at 00, 12 and 18 hours
at various stations in regions A, B and C. Stations are numbered as in Fig.~\ref{fig2}.
For each station, the first triplet of bars denotes the wet season (00, 12, 18 hours), the
second triplet denotes the dry season.
Dots in the upper (positive) and lower (negative) part of the diagram indicate, respectively, $\Delta P$
and $\Delta p$ values that differ significantly from zero at 0.01 probability level (Student $t$-test).
E.g. the first pair of brown and green bars for station 1 (Piren\'{o}polis) in region A indicates that
on those days when the WNW winds blow at 00 hours
the daily rainfall is on average 2 mm~d$^{-1}$ higher, while surface pressure at 00 hours is 0.5 hPa lower,
than on those days when at 00 hours the winds blow from the ESE direction. Both differences are statistically significant (two dots).
See Table 2 in the Supplementary Information for
data for each station.
}
\end{figure*}

\begin{figure*}[h]
\centerline{
\includegraphics[width=0.99\textwidth,angle=0,clip]{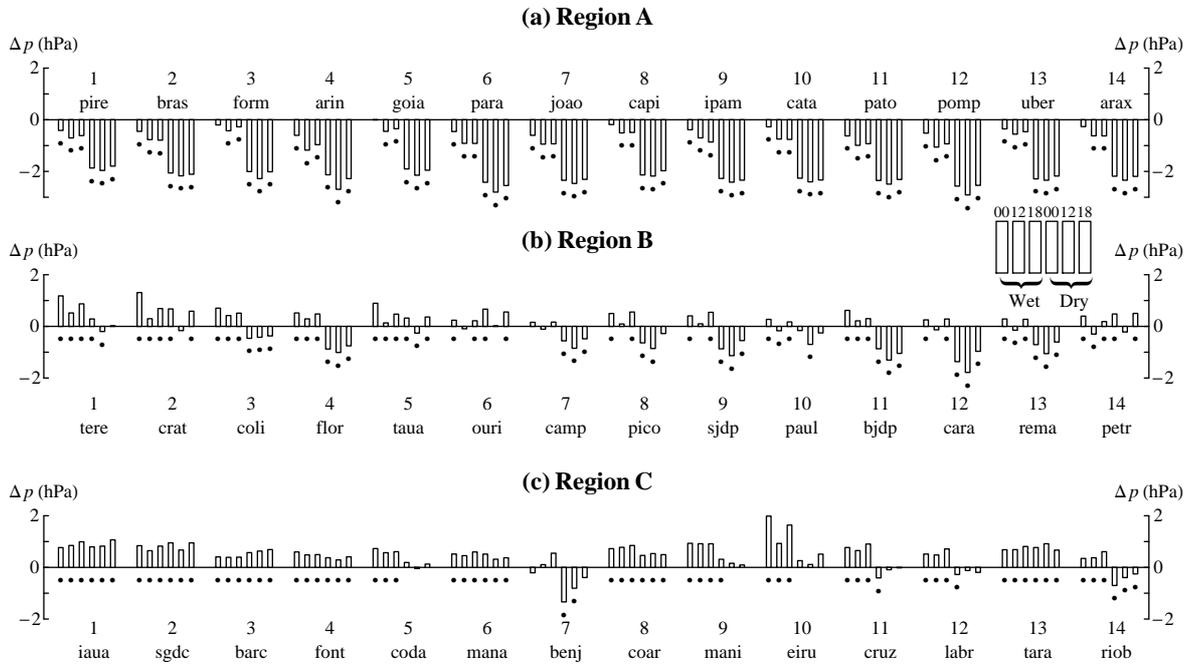}
}
\caption{\label{fig4}
Difference in mean pressure between rainy and rainless days at different times of the day (00, 12 and 18) on different stations during wet and dry season
in regions A, B, C. Stations are numbered as in Fig.~\ref{fig2}. For each station, the first triplet of bars denotes the wet season, the
second triplet denotes the dry season. Dots indicate those mean differences that differ significantly from zero at 0.01 probability level (Student $t$-test).
See Table 2 in the Supplementary Information for numerical data.
}
\end{figure*}

\begin{figure*}[h]
\centerline{
\includegraphics[width=0.5\textwidth,angle=0,clip]{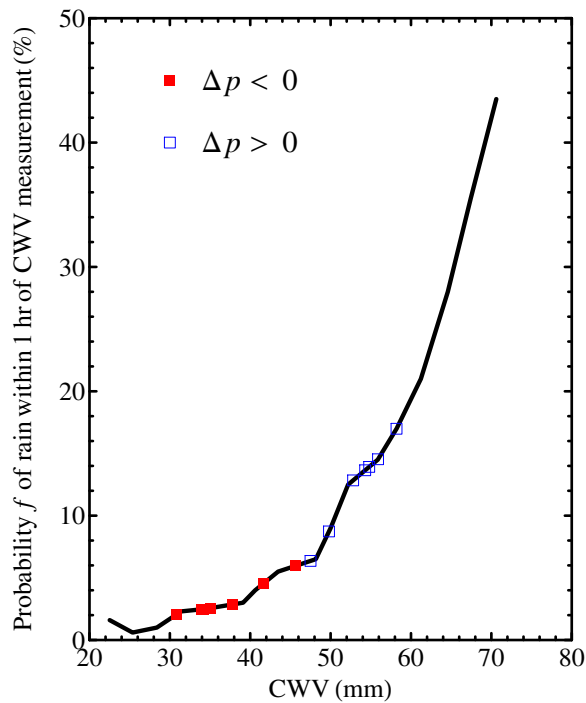}
}
\caption{\label{fig5}
Probability of rain as dependent on CWV. Solid curve $f({\rm CWV})$: data of Fig.~10b of \citet{holloway10}
with separate points in the original graph joined by straight lines. Boxes: $f({\rm CWV^+})$ where CWV$^+$ values
represent mean CWV during rainy days for those lines in Table 2 where $\Delta p < 0$
(red filled boxes) or $\Delta p > 0$ (empty blue boxes).
}
\end{figure*}

\begin{figure*}[h]
\centerline{
\includegraphics[width=0.5\textwidth,angle=0,clip]{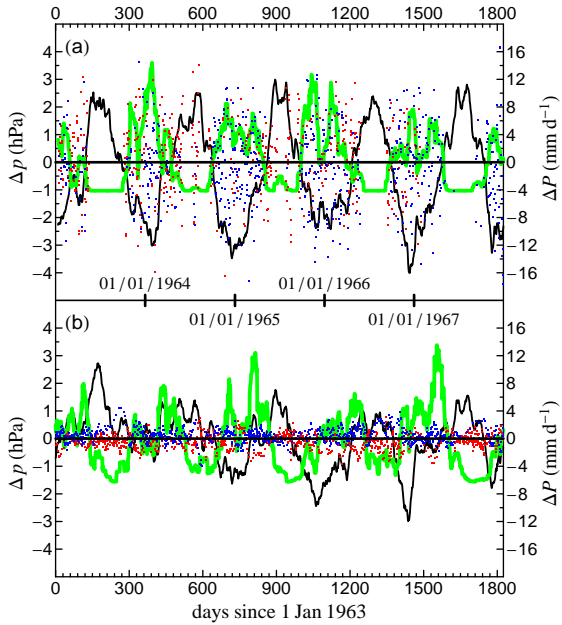}
}
\caption{\label{fig6}
Five years (1963-1967) of changes in pressure (thin black line) and rainfall (thick green line) in (a) Brasilia (station A2) and (b) Manaus (station C6).
Pressure (12 hr measurement) and rainfall (daily measurement) are smoothed by moving average over $m = 31$~d in Brasilia and $m = 7$~d
in Manaus. The long-term mean values are subtracted (888.1~hPa and 4.1~mm~d$^{-1}$ in Brasilia and 1005.8~hPa and 6.3~mm~d$^{-1}$ in Manaus).
In (a), daily pressure minus smoothed mean ($\Delta p_i$) is shown for individual rainy days (blue boxes) and rainless days (red boxes).
Only those days are considered that have at least one rainy and at least one rainless day within the spell of $m$ days centered at a given day
(note lack of boxes during the dry season in Brasilia). In (b), the $\Delta p_i$ values are additionally smoothed by moving average
over $m=7$~d to show the predominantly positive (negative) location of blue (red) boxes.
}
\end{figure*}

\begin{figure*}[h]
\centerline{
\includegraphics[width=0.9\textwidth,angle=0,clip]{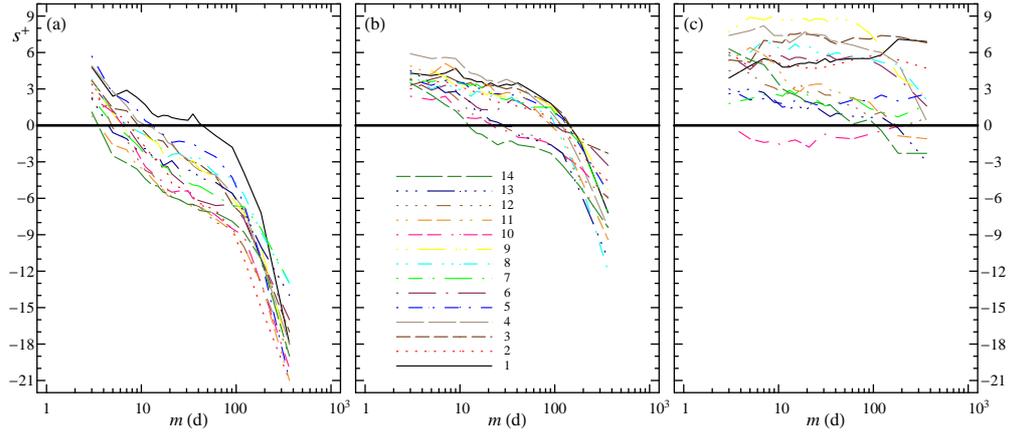}
}
\caption{\label{fig7}
Prevalence of high pressure during rainy days as dependent on the length $m$ (d) of the smoothing interval
in (a) region A, (b) region B and (c) region C.
The value of $s^+(m)\equiv (n^+ - N^+/2)/\sqrt{N^+/2}$ is the statistical significance of the effect,
where $n^+$ is the number of rainy days during which pressure is higher than the mean pressure of $m$
consecutive days centered at a given day and containing at least one rainless day, $N^+$ is the total
number of rainy days considered, and $\sigma = \sqrt{N^+/2}$ is the standard deviation of $n^+$
assuming equal probability of rainy days having higher or lower pressure than the smoothed average.
Negative $s^+$ values indicate that rainy days have lower pressure. $|s^+| > 3$ indicates statistical significance
of the effect at $>3\sigma$. The following values of $m$ are analyzed:
$m =$ 3, 5, 7, 9, 11, 13, 17, 19, 21, 25, 31, 35, 41, 61, 91, 121, 183 and 365. Stations are numbered as in Fig.~\ref{fig2}.
}
\end{figure*}

Here using independent evidence (long-term wind, rainfall and pressure records) for 28 meteorological stations in two non-forested regions in Brazil
(Fig.~\ref{fig2}) we showed that the winds blowing from the Amazon rainforest are indeed associated with higher rainfall and a higher column water vapor content
(CWV) than winds coming to the same station from non-forest regions  (Fig.~\ref{fig3} and Table~2).
However, the CWV values observed in these regions do not exceed 50 mm. Such relatively low CWV values have a low probability of triggering
intense convection and rainfall (Fig.~\ref{fig5}). Rather, rainfall appears to be imported to the non-forested regions by the low pressure systems
that originated elsewhere, in particular, over the Amazon forest. This is consistent with our finding that rainy days in regions A and B
often have lower pressure than rainless days especially during the dry season when the CWV content is low (Fig.~\ref{fig4}a,b,
Table~2, Fig.~\ref{fig7}a,b).
At the same time, our analysis revealed interesting differences in the rainfall-associated atmospheric pressure between
the forest (C) and non-forest (A, B) regions. In the forest region rainy days are associated with higher atmospheric pressure than are
rainless days (Figs.~\ref{fig4}c and~\ref{fig7}c). We explained how this difference
may indicate the more spatially and temporarily uniform rainfall in the Amazon forest compared to the non-forest areas.

The tall trees of the rainforest ensure high canopy roughness. This roughness suppresses strong winds
in the lower atmosphere that harbors most moisture. Lack of strong winds makes the moisture content in the atmospheric column above the forest canopy
less influenced by the atmospheric processes outside the forest.
Another salient feature of rainforests -- very high leaf area index -- allows for high intensity of evaporation from
the forest canopy \citep[e.g.,][]{calder86,fisher09}. The characteristic mean CWV values maintained in the forest region during both wet and dry season are
near or above 50 mm (Table~2). This is the region of CWV values where the probability of tropical rainfall has a highest sensitivity
to CWV: the probability of precipitation within an hour of observed CWV value increases at a rate of over 1\%/mm CWV.
For comparison, at CWV$\le 40$~mm  this rate is an order of magnitude lower (Fig.~ \ref{fig5}).
The high value of mean CWV in the rainforest makes it possible for the local vegetation cover to trigger or suppress convection and rainfall
by small changes in conditions that can be achieved by transpiration, emission of biogenic condensation nuclei and other biotically
mediated processes. Stable maintenance of high condensation intensity over the forest leads to formation of a large-scale low pressure zone
and facilitates the needed import of atmospheric moisture from the ocean.
Thus the rainforest emerges as a most complex self-sustainable rainmaking system on land,which is characterized by both unique intensity
as well as by remarkable spatial and temporal uniformity \citep{millan11,mgl13}.
Vegetation in the dry regions should be generally unable to trigger precipitation in this way and becomes more passive recipients of low pressure
rain-yielding systems formed elsewhere \citep[but see][]{ch10}.

\begin{figure*}[h]
\centerline{
\includegraphics[width=0.5\textwidth,angle=0,clip]{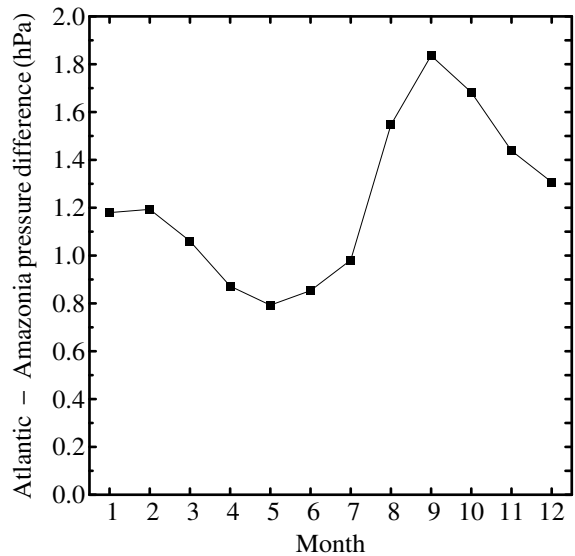}
}
\caption{\label{fig8}
Long-term mean monthly sea level pressure difference between the Atlantic
(equator-10$^{\rm o}$N, 50$^{\rm o}$-30$^{\rm o}$W) and Amazonia
(10$^{\rm o}$S-equator, 70$^{\rm o}$-50$^{\rm o}$W). Data (1960-2012) taken from
NCAR-NCEP re-analysis \citep{kalnay96}.
}
\end{figure*}

Recognition that the Amazon forest determines regional atmospheric circulation helps suggest how deforestation might impact the rainfall in non-forested regions (like A and B).  The prevailing
winds that bring most annual rainfall to regions A and B come from easterly directions (E, SE, NE) (Fig.~2). These winds are facilitated by
the stable low pressure zone over the Amazon rainforest and the generally higher pressure over the Atlantic ocean.
The horizontal pressure gradient induced by condensation was theoretically estimated as $-\pt p/\pt x = (p_v/h_\gamma)(w/u) [1 - (h_v/h_\gamma) (u/w) \alpha]$
\citep[][Eq.~12]{m13,mg10}, where $p_v$ is water vapor partial pressure, $h_\gamma$ is the exponential scale height of the water vapor mixing ratio
$\gamma \equiv p_v/p$, $h_v$ is the exponential scale height of $p_v$, $w$ and $u$ are vertical and horizontal velocities, respectively,
and $\alpha \equiv -(\pt T/\pt x )/(\pt T/\pt z)$ is the ratio of the horizontal to vertical temperature gradients taken with the minus sign.
All parameters entering this estimate are directly observable. According to \citet[][Fig.~4]{arraut12}, the mean temperature difference
in the lower 1.5~km between the Atlantic (equator-10$^{\rm o}$N, 50$^{\rm o}$-30$^{\rm o}$W) and Amazonia
(10$^{\rm o}$S-equator, 70$^{\rm o}$-50$^{\rm o}$W) is approximately 2$^{\rm o}$C per $2\times 10^3$~km, such that we can estimate $\pt T/\pt x = 10^{-3}$~
$^{\rm o}$C~km$^{-1}$. Taking $w = 0.3$~mm~s$^{-1}$, $u = 5$~m~s$^{-1}$ in the lower 1.5~km \citep[][Figs.~4,~12]{zhou98},
$p_v$ = 30~hPa, $h_\gamma = 9$~km, $h_v = 4.5$~km and the moist adiabatic lapse rate of
$-\pt T/\pt z = 4.5$~$^{\rm o}$C~km$^{-1}$ \citep{m13,jas13}, we obtain $-\pt p/\pt x = 6.6\times10^{-4}$~hPa~km$^{-1}$. This corresponds
to a pressure difference of 1.3~hPa over a distance of 2000~km. The mean monthly sea level pressure difference between the Atlantic (higher pressure) and
Amazonia (lower pressure) changes from 0.8 to 1.8 hPa and approximately agrees with the theoretical estimate (Fig.~\ref{fig8}).

The large-scale pressure gradients that drive the condensation-induced air motion are proportional to the intensity of local
condensation and, hence, precipitation \citep{m13}. In the stationary case $P = {\rm ET} + C$,
where $C$ is the net amount of atmospheric moisture imported to the region.
Rather than merely influencing the moisture content in the air that is passing over the forest, the process of evapotranspiration
can impact regional atmospheric dynamics by enhancing rainfall and thus modifying the large-scale pressure gradients.
This, in turn, enhances and stabilizes precipitation in a positive feedback loop.
If deforestation is accompanied
by erosion of the regional-scale low pressure zone as we predict, ocean-to-land winds and  rainfall will decline. Conversely, restoration of forests will both increase local rainfall and also  contribute to
strengthening of the total continental ocean-to-land moisture transport and associated feedbacks, increasing both the magnitude and reliability of
rainfall in a wider region.

\noindent
{\bf Acknowledgements.} We are grateful to D.~V.~Spracklen, S.~R.~Arnold and C.~M.~Taylor for constructive attention to our comments. We thank
our reviewers for constructive criticisms and suggestions. The radiosonde data were provided by the University of Wyoming:
we gratefully acknowledge the support of Larry Oolman. BLL thanks the University of California Agricultural Experiment Station for their
partial support. Supplementary Information is available at \url{http://www.bioticregulation.ru/common/pdf/spr/sprn-sup.pdf}.

\bibliographystyle{ametsoc}
\bibliography{met-refs}

\newpage

\newpage

\newpage

\newpage

\newpage

\newpage

\end{document}